\documentclass[final,5p,times,twocolumn]{elsarticle}

\usepackage{amssymb}
\usepackage{amsthm}
\usepackage{amsmath}

\journal{Physica D}

\begin{document}
 
\begin{frontmatter}



\title{Determining the source of phase noise: Response of a driven Duffing oscillator to low-frequency damping and resonance frequency fluctuations}


\author{C.\,S. Barquist\corref{cor1}}
\ead{cbarquist@ufl.edu}
\author{W.\,G. Jiang}
\author{K. Gunther}
\author{Y. Lee\corref{cor1}}
\ead{ysl@ufl.edu}
\cortext[cor1]{Corresponding Authors}

\address{Department of Physics, University of Florida, Gainesville, Florida, 32611, USA}

\begin{abstract}
We present an analytical calculation of the response of a driven Duffing oscillator to low-frequency fluctuations in the resonance frequency and damping. We find that fluctuations in these parameters manifest themselves distinctively, allowing them to be distinguished. In the strongly nonlinear regime, amplitude and phase noise due to resonance frequency fluctuations and amplitude noise due to damping fluctuations are strongly attenuated, while the transduction of damping fluctuations into phase noise remains of order $1$. We show that this can be seen by comparing the relative strengths of the amplitude fluctuations to the fluctuations in the quadrature components, and suggest that this provides a means to determine the source of low-frequency noise in a driven Duffing oscillator.

\end{abstract}


%%\item Research highlight 1
%%\item Research highlight 2


\end{frontmatter}


\section{Introduction}
\label{section:introduction}

Phase noise (also frequency noise) is ubiquitous in the field of oscillators. Understanding the fundamental processes that cause this noise is of interest to both physicists and engineers \cite{Hajimiri1998,lee2000oscillator,Sansa2016}. From the engineer's point of view, a clear understanding of the fundamental noise processes leads to better designed oscillators, and more stable oscillator frequencies. Improved oscillator stability leads to more accurate time keeping, more precise GPS, more accurate Doppler radar, increased data transmission rates, reduced transmission power, {\it etc.} \cite{Leeson2016}. In the world of micro- and nano- mechanical oscillators, increased frequency stability has facilitated high resolution neutral mass spectrometry, with the eventual goal of reaching single Dalton resolution \cite{Ekinci2004,Ekinci2005}. From the physicist's point of view, noise carries important information about a system under study, from gaining access to the spectrum of fundamental excitations through the fluctuation dissipation theorem \cite{Kubo1966}, to constraining models of the early universe from observations of the noisy cosmic microwave background \cite{Samtleben2007} .

In this paper we are concerned with the question of how phase noise can be transduced in an open-loop driven Duffing oscillator, and whether it is possible to determine the method of transduction from measurements of the noise. This question initially arose from recent experiments involving micro-electromechanical (MEMS) oscillators interacting with quantum turbulence in superfluid $^4$He \cite{Barquist2020,Barquist2019}. In these experiments, when the MEMS was operated in the strongly nonlinear regime, it was observed that turbulent fluctuations were being transduced into phase noise. In the context of these experiments, it is known that the damping of the device is significantly affected by the presence of quantum vortices \cite{Barquist2020}, and the density of vortices around the device fluctuates. Therefore, it is expected that turbulent fluctuations lead to damping fluctuations, which then induce phase noise. However, we have found that there is minimal discussion on damping induced noise in driven Duffing oscillators. Recently, measurements of a SiN  NEMS resonator have demonstrated a method for measuring damping fluctuations in a driven Duffing oscillator \cite{Maillet2018}. However, that method requires many measurements of the transition frequencies in order to determine their statistical distribution. For devices with long relaxation times, this method can be time consuming. The transduction of parameter noise into phase noise for nonlinear resonators has also been discussed in the context of optimal operating points \cite{Kenig2012,Villanueva2013,Agarwal2006,Brenes2016,Agrawal_2014}. There, the discussion is focused on feedback driven (closed-loop) configurations and any discussion of damping fluctuations is cursory. Here, we present a complementary analysis of the transduction of parameter noise into phase noise for open-loop driven nonlinear resonators, with emphasis on damping fluctuations. We also describe a means to discriminate between damping and other parameter fluctuations which does not rely on lengthy measurements.

\begin{figure*}
	\centering
	\includegraphics[width=\linewidth]{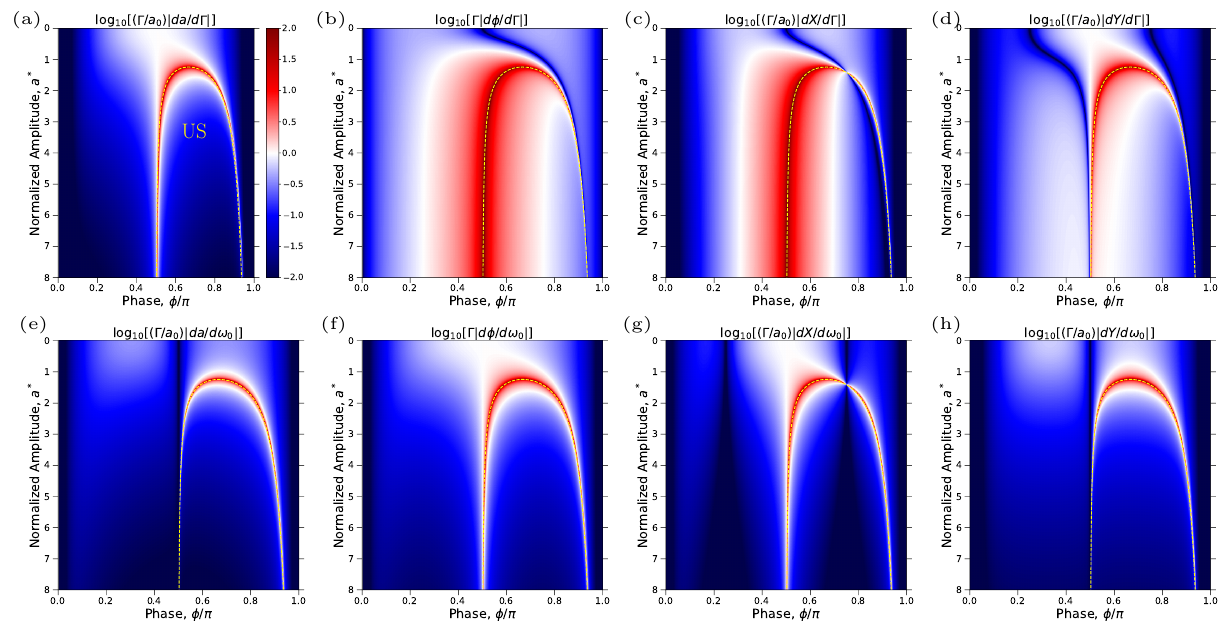}
	\caption{Response functions for an incremental change in either the damping, $\Gamma$, or the resonance frequency, $\omega_0$, plotted as a function of both phase of the oscillator, $\phi$, and the normalized amplitude, $a^*$. The response functions plotted in log scale and normalized by $\Gamma$ and $a_0$ where appropriate to make them dimensionless. (a)--(d) Response to changes in $\Gamma$ for $a$, $\phi$, $X$, and $Y$, respectively. (e)--(h) Response to changes in $\omega_0$ for $a$, $\phi$, $X$, and $Y$, respectively. The color scale indicated in (a) is representative of all plots. The border between stable and unstable oscillations ({\it i.e.} $d\omega/da = 0$) is shown in all plots by a dashed yellow line. The dome of the unstable region is labeled in (a) as ``US".}
	\label{fig:bare_response}
\end{figure*}

For mechanical oscillators, sources of noise can be extrinsic -- from the coupling of the oscillator to the driving and measurement circuitry -- or intrinsic -- from unavoidable coupling of the oscillator to environment. An example of extrinsic noise is thermal Johnson noise from resistive elements in the measurement circuit \cite{Ekinci2005}, while some examples of intrinsic noise are thermomechanical \cite{Ekinci2004,Ekinci2005,Cleland2002}, thermoelastic \cite{Ekinci2004,Ekinci2005,Cleland2002}, adsorption-desportion \cite{Ekinci2005,Cleland2002,Dykman2010}, diffusion noise \cite{Cleland2002,Atalaya2011}, and defect motion \cite{Fong2012}. These processes can be further classified into additive noise, where the noise results from a random force in addition to the drive, and multiplicative (or parametric) noise, where the noise arises from a fluctuating device parameter, such as the mass. A thorough discussion of these noise processes and their effects on a nano-mechanical oscillator is can be found in Refs. \cite{Vig1999,Cleland2002}.

The canonical additive noise for a mechanical oscillator is thermomechanical fluctuations. These fluctuations arise from the device being in thermal equilibrium with its surroundings, and result in a random force with a white spectrum acting on the device. This force results in random motion of the device. When this motion is superimposed on the driven motion of the device, it appears as a noise in the amplitude and phase of the oscillator. This type of noise is always associated with energy loss and can be directly related to the damping of the oscillator, $\Gamma$, or the inverse of the quality factor, $1/Q$, through the fluctuation dissipation theorem.

Multiplicative, or parametric noise, differs from the additive noise in that the parameters of the oscillator are changing with time. Changes in the oscillator parameters then lead to changes in the phase or amplitude of oscillation. For small parametric fluctuations, the power spectrum of the observed noise is related to the power spectrum of the fluctuating parameter \cite{Cleland2002} through 
\begin{equation}
S_{\alpha}(\omega) = \left|\frac{d\alpha}{d\beta}\right|^2S_{\beta}(\omega),
\label{eq:spectral_relation}
\end{equation}
where 
\begin{equation}
S_{\alpha}(\omega) = \int_{-\infty}^{\infty}\alpha(t)\alpha^*(t)e^{-i\omega t}dt,
\label{spectral_definition}
\end{equation}
and $\alpha$ and $\beta$ are indices representing the fluctuating parameters, {\it e.g.} $\alpha = \phi$ (phase) and $\beta = m$ (mass). In the literature, it is commonly considered that parametric phase noise is mediated through fluctuations of the resonance frequency, $\omega_0 = \sqrt{k/m}$, which is true for most operating environments \cite{Sansa2016}. Unless the resonator is in a perfect vacuum, the surface of the oscillator will be bombarded with gaseous particles that are constantly adsorbing and desorbing from the surface. If the resonator's mass is sufficiently small, the fluctuating mass of adsorbed particles will cause an appreciable change in the resonance frequency of the device \cite{Ekinci2004,Ekinci2005}. Changes in the stiffness will also shift the resonance, which can occur from fluctuations of the temperature of the oscillator, through the temperature dependent elastic modulus and internal stress \cite{Sansa2016,Ekinci2004}. Interestingly, to date none of these mechanisms can explain the excess phase noise observed in N/MEMS resonators \cite{Sansa2016}.

While it may be uncommon, it is also possible to have fluctuations in the damping rate, $\Gamma$. These fluctuations often negligibly affect the oscillator response, and are ignored. For example, take the case of a fluctuating mass. Fluctuations in the mass will also cause fluctuations in $\Gamma$, because $\Gamma = \gamma/2m$, where $\gamma $ is the constant of proportionality for some linear damping force, {\it i.e.} $F = \gamma v$. However, these fluctuations are suppressed by a factor of $1/Q$ compared to fluctuations in $\omega_0$. The typical method of operating the resonator exactly on resonance in the linear regime also renders the damping fluctuations unimportant. In this regime of operation, $d\phi/d\Gamma = 0$, and fluctuations in the damping are not transduced into fluctuations of the phase. However, for certain applications, when large amplitudes of motion are needed, the nonlinear behavior may become important. When operating in the nonlinear regime, we must revisit the response of the oscillator to parametric fluctuations and determine whether damping fluctuations become important. If they are important, can their effect on the phase noise be distinguished from fluctuations in $\omega_0$?
\begin{figure}
	\centering
	\includegraphics[width=0.9\linewidth]{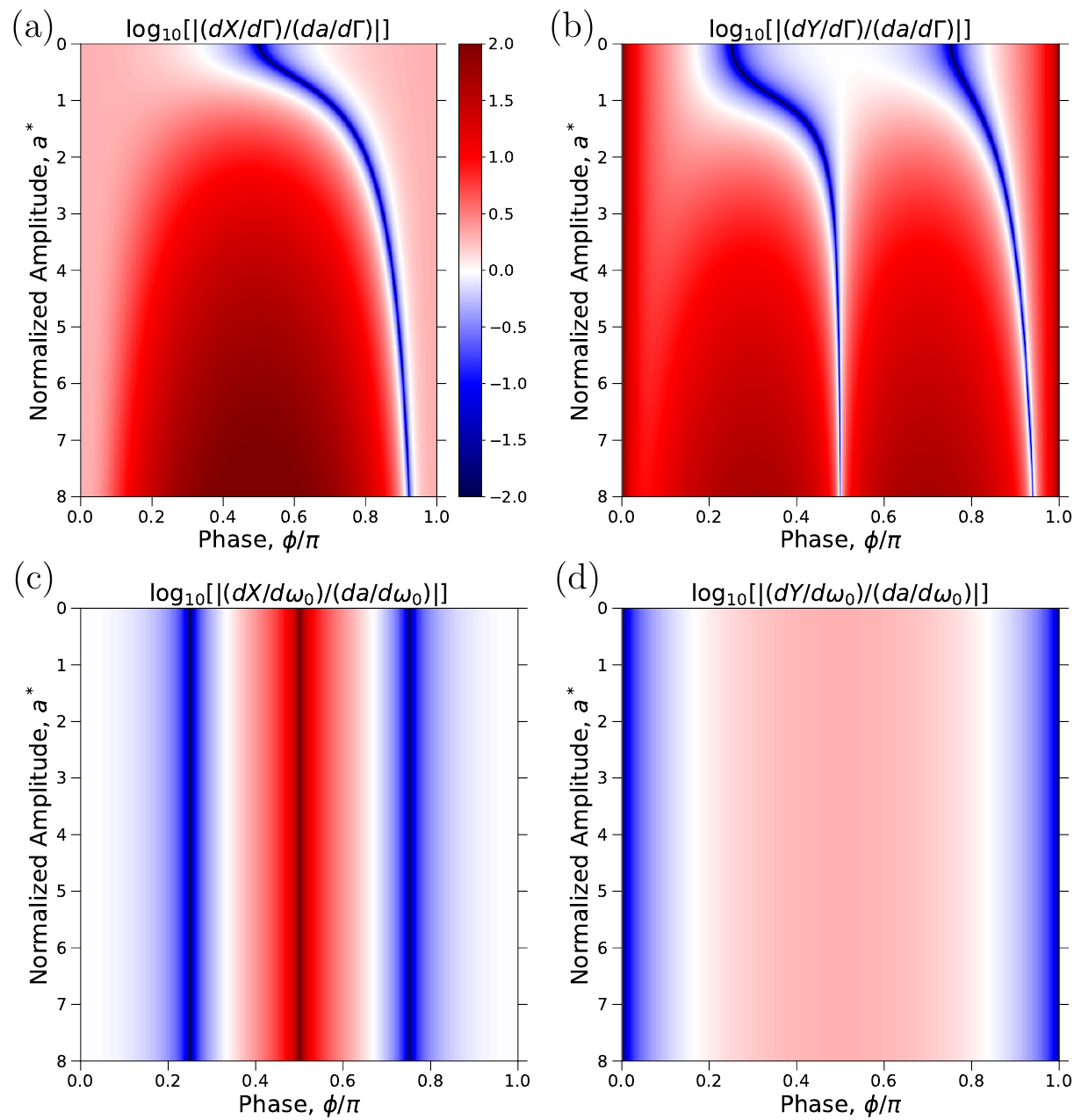}
	\caption{Responses of the quadrature components relative to the amplitude response for changes in damping, $\Gamma$, or resonance frequency $\omega_0$, plotted as a function of both the phase of the oscillator, $\phi$, and the normalized amplitude, $a^*$. In the strongly nonlinear regime, $(a^*)^2 \gg 1$, resonance frequency fluctuations can be distinguished from damping fluctuations by the relative strength of the quadrature and amplitude responses off resonance ({\it i.e.} $\phi \neq 2\pi$). The color scale indicated in (a) is representative for all plots. }
	\label{fig:scaled_response}
\end{figure}

To answer these questions, we consider the first order response of a driven Duffing oscillator to small low-frequency fluctuations in $\omega_0$ and $\Gamma$, and compare the responses of the oscillator. The fluctuation frequency is considered low if the time scale of the fluctuations, $\tau$, is larger than the response time of the oscillator, $1/\Gamma$, {\it i.e.} $\tau > 1/\Gamma$. In this limit, $\omega_0$ and $\Gamma$ are stationary over the response time of the oscillator, and any transient response from the changing parameters is damped and negligible. To determine the response of the oscillator to fluctuations of this type, it is sufficient to determine the steady state response for constant values of $\omega_0$ and $\Gamma$, and introduce the time dependence of these parameters directly into the steady state solution.

We consider an oscillator driven in an open-loop configuration by a noiseless ideal-source, {\it i.e.,} constant driving frequency, $\omega$, and constant driving force amplitude, $f_0$. This system is described by the following differential equation:
\begin{equation}
\ddot{x} + 2\Gamma\dot{x} + \omega_0^2 x + \alpha_3 x^3 = g_0 \cos(\omega t),
\label{eq:eom}
\end{equation}
where $\alpha_3$ characterizes the conservative nonlinear restoring force and $g_0 = f_0/m$. A general solution to Eq.\,\ref{eq:eom} will have the form $x(t) = a(t)\cos(\omega t +\phi(t))$, where $a(t)$ and $\phi(t)$ vary slowly over one period of motion. The steady state solution is determined by setting $\dot{a} = a\dot{\phi} = 0$.

The solution to Eq.\,\ref{eq:eom}, with static parameters and small $\alpha_3$, can be found in most textbooks on nonlinear oscillators, {\it e.g.} Refs.\,\cite{Nayfeh1995,L.D.Landau1995,IvanaKovacic2011}. Upon solving, the following relationships are found:
\begin{align}
g_0\sin\phi &= 2a\Gamma\omega \label{eq:eom-sin},\\
g_0 \cos\phi &= a\left(\omega_0^2 - \omega^2 + 2\Pi a^2\omega_0\right) \label{eq:eom-cos}.
\end{align} 
Here, $\Pi = \frac38 \frac{\alpha_3}{\omega_0}$ characterizes the amplitude dependent resonance frequency shift. These equations can be rearranged to yield the common description of the Duffing oscillator
\begin{align}
&a(\omega) = \frac{g_0}{\sqrt{\left(\omega_0^2 - \omega^2 + 2\Pi a^2 \omega_0 \right)^2 + \left(2\Gamma \omega\right)^2}}, \\
&\tan(\phi(\omega)) = \frac{2\Gamma\omega}{\omega_0^2 - \omega^2 + 2\Pi a^2 \omega_0}.
\end{align}
From the experimental point of view, it is common to work with the quadrature components:
\begin{align}
X &= a\cos\phi, \label{eq:quad-x}\\
Y &= a \sin\phi. \label{eq:quad-y}
\end{align}
Typically, $X$ and $Y$ are referred to as the {\it in-} and {\it out-of-phase} components, respectively.

Determining the spectrum of fluctuations of the dynamical variables ($a$ and $\phi$ or $X$ and $Y$) from the spectrum of the oscillator parameters ($\omega_0$ and $\Gamma$) is equivalent to computing the first derivatives of the dynamical variables with respect to the oscillator parameters, as indicated by Eq.\,\ref{eq:spectral_relation}.

\section{Results}

Here, we present the analytical solutions and their plots for varying amplitudes and phases of oscillations. We then provide a discussion of the difference in responses and identify how to discriminate phase noise driven by resonance frequency fluctuations from phase noise driven by damping fluctuations.

\paragraph{Damping Fluctuations}

The amplitude and phase responses are found by implicitly differentiating Eqs.\,\ref{eq:eom-sin}\,and\,\ref{eq:eom-cos}, then solving the system of equations. When computing the response to changes in damping, we take $g_0$, $\Pi$, $\omega_0$, and $\omega$ to be constant. This yields the following response functions:
\begin{align}
\Gamma \frac{d\phi}{d\Gamma} &= \frac{\cos\phi\sin\phi + 2(a^*)^2\sin^4\phi }{1 + 2 (a^*)^2 \cos\phi\sin^3\phi}, \label{eq:dphi_dgam}\\
\Gamma \frac{da}{d\Gamma} &= \frac{-a_0\sin^3\phi}{1 + 2 (a^*)^2 \cos\phi\sin^3\phi}. \label{eq:da_dgam}
\end{align}
Here $a_0 = g_0/2\Gamma \omega_0$ is the amplitude of motion on resonance ({\it i.e.} $\phi = \pi/2$), and $(a^*)^2 = a_0^2\Pi/\Gamma$ is dimensionless parameter which measures the degree of nonlinearity. In the linear regime $(a^*)^2 = 0$. 

The quadrature response is determined by differentiating Eqs.\,\ref{eq:quad-x}\,and\,\ref{eq:quad-y}, which yields
\begin{align}
\Gamma\frac{dX}{d\Gamma} &= a_0 \left[\left(2\cos^2\phi -1\right)\left(\Gamma\frac{d\phi}{d\Gamma}\right) - \cos\phi\sin\phi\right], \label{eq:dx_dgam}\\
\Gamma\frac{dY}{d\Gamma} &= a_0 \left[2\sin\phi\cos\phi \left(\Gamma\frac{d\phi}{d\Gamma}\right) - \sin^2\phi\right]. \label{eq:dy_dgam}
\end{align}
It is notable that the quadrature response is directly related to the phase response, $d\phi/d\Gamma$. By comparing the quadrature fluctuations to the amplitude fluctuations, we have a measure of the relative strength of the phase and amplitude fluctuations. The solutions to Eqs.\,\ref{eq:dphi_dgam}--\ref{eq:dy_dgam} for $0 \leq a^* \leq 8$ and $0 \leq \phi \leq \pi$ are shown in Fig.\ref{fig:bare_response}\,(a)--(d) in log scale.

\paragraph{Frequency Fluctuations}
The response to resonance frequency fluctuations is calculated in the same way as before, except that $g_0$, $\Pi$, $\Gamma$, and $\omega$ are held constant. The following response functions are found:

\begin{align}
\Gamma\frac{d\phi}{d\omega_0} &= \frac{-\sin^2\phi}{1+2(a^*)^2\sin^3\phi\cos\phi} \label{eq:dphi_domg}\\
\Gamma\frac{da}{d\omega_0} &= \frac{-a_0\cos\phi\sin^2\phi}{1+2(a^*)^2\sin^3\phi\cos\phi} \label{eq:da_domg} \\
\Gamma\frac{dX}{d\omega_0} &= a_0\left(2\cos^2\phi - 1\right)\left(\Gamma\frac{d\phi}{d\omega_0}\right)  \label{eq:dx_domg} \\
\Gamma\frac{dY}{d\omega_0} &=  2a_0\sin\phi\cos\phi\left(\Gamma\frac{d\phi}{d\omega_0}\right) \label{eq:dy_domg}
\end{align}
Similarly, the quadrature response is directly related to only the phase response, $d\phi/d\omega_0$. The solutions to Eqs.\,\ref{eq:dphi_domg}--\ref{eq:dy_domg} for $0 \leq a^* \leq 8$ and $0 \leq \phi \leq \pi$ are shown in Fig.\ref{fig:bare_response}\,(e)--(h) in log scale. 

\paragraph{Discussion}

From Fig.\,\ref{fig:bare_response}, we observe several distinctive features between damping and resonance frequency driven fluctuations. Firstly, when the fluctuations are present in $\omega_0$, the response of the device on resonance ($\phi = \pi/2$), is independent of the amplitude of motion, even in the nonlinear regime, see Fig\,\ref{fig:bare_response}(e)-(h). For fluctuations in $\Gamma$, this is only true for the amplitude response (Fig.\,\ref{fig:bare_response}(a)), while the phase response (Fig.\,\ref{fig:bare_response}(b)) goes from zero in the linear regime to a large value in the nonlinear regime, which scales as $(a^*)^2$, see Eq.\,\ref{eq:dphi_dgam}. Because of this behavior, fluctuations in $\Gamma$ can no longer be ignored when operating in the nonlinear regime. Secondly, in the strongly nonlinear regime, $(a^*)^2 \gg 1$, the response functions are peaked where the solutions transition between the stable and unstable branches. The border between the stable and unstable regions is located where $d\omega/da = 0$, and is shown in Fig.\,\ref{fig:bare_response} as a dashed yellow curve. Away from these points the responses are attenuated. While all responses are attenuated away from the transition points, the attenuation in the phase channel from damping fluctuations (Fig.\,\ref{fig:bare_response}(b)) is more gradual, especially when in comparison to the attenuation of the amplitude response function (Fig.\,\ref{fig:bare_response}(a)). This gradual attenuation is mirrored in the response of the quadrature channels (Fig.\,\ref{fig:bare_response}(c)-(d)), whose responses are determined by the phase response. 

\begin{figure}
	\centering
	\includegraphics[width=\linewidth]{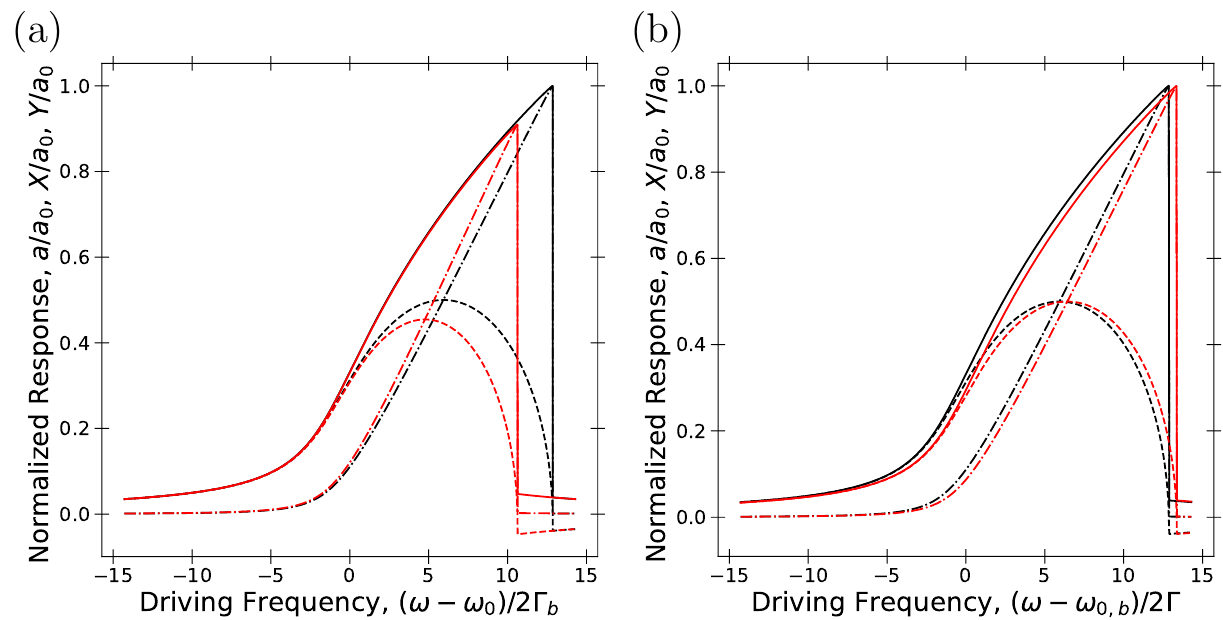}
	\caption{Stable frequency responses of a Duffing oscillator for a upward frequency sweep. The solid, dashed, and dot-dashed curves represent the $a$, $X$, and $Y$ channels, respectively. (a) Responses for two different damping parameters. Here $\Gamma_r = 1.1\Gamma_b$, while all other parameters are fixed. Curves $r$ and $b$ are shown in red and black, respectively. (b) Responses for two different frequency parameters. Here $\omega_r-\omega_b = \Gamma$, while all other parameters are fixed. For both figures the curves correspond to $a^* = 5.1$.}
	\label{fig:gam_omg_comp}
\end{figure}

The differing attenuation in the responses suggests a simple method for discriminating between the origin of noise in the driven Duffing system: comparison of the relative strengths of the amplitude and quadrature noise off resonance. Figure\,\ref{fig:scaled_response} shows the response of the quadrature channels scaled by the response of the amplitude channel for both damping and resonance frequency fluctuations. It can be seen that away from resonance and the transition points, the quadrature response is significantly larger than the amplitude response, when damping fluctuations are dominant. As the device is driven further into the nonlinear regime this discrepancy increases. This effect can be noticed in the scaling of response functions with $a^*$, off resonance. For large $a^*$, $da/d\Gamma \sim 1/(a^*)^2$ while $d\phi/d\Gamma$, $dX/d\Gamma$, and $dY/d\Gamma$ all become independent of $a^*$ and of order $1$. While, for resonance frequency driven fluctuations, all response functions scale as $1/(a^*)^2$. Therefore, driving the Duffing oscillator into the strongly nonlinear regime acts to filter out noise except for damping driven phase noise, and actually amplifies the phase noise when driven on resonance. Indeed, very similar behavior has been observed in a SiN NEMS oscillator driven into the nonlinear regime \cite{Maillet2018}.

To better understand how the damping and resonance frequency changes result in changes of the amplitude and phase, it is helpful to consider the frequency response of the oscillator. Several frequency responses of the Duffing oscillator are shown in Fig.\,\ref{fig:gam_omg_comp}. Only the stable branches are shown for an upward frequency sweep. The solid, dashed, and dot-dashed curves correspond to the amplitude ($a$), in-phase ($X$), and out-of-phase ($Y$) channels, respectively. Figure\,\ref{fig:gam_omg_comp}(a) shows the response of the Duffing oscillator with different damping parameters. Curve $r$ (red) has $10\%$ larger damping than curve $b$ (black), $\Gamma_r=1.1\Gamma_b$. For constant driving frequency, it can be seen that the quadrature components vary significantly compared to the amplitude, when off resonance. In Fig.\,\ref{fig:gam_omg_comp}(b) the frequency response of the Duffing oscillator with two different resonance frequencies is shown. The resonance frequencies of the curves differ by $\Gamma$, {\it i.e.,} $\omega_r-\omega_b = \Gamma$. Here, it can be seen that the changes in the quadrature components are comparable to the change in the amplitude, almost everywhere.

\section{Conclusion}
With the goal of determining the importance of $\Gamma$ fluctuations on the nonlinear Duffing oscillator, we calculated the response of the oscillator to low-frequency fluctuations in $\Gamma$ and $\omega_0$.  We demonstrated that on resonance, the response of the oscillator to fluctuations in $\omega_0$ is unaltered in the nonlinear regime. In contrast, the response to fluctuations in $\Gamma$ becomes important, and are amplified by driving further into the nonlinear regime. We also demonstrated that in the strongly nonlinear regime,  $(a^*)^2 \gg 1$, the source of noise can be determined by comparing the relative strength of the amplitude and quadrature fluctuations off resonance.
\section{Acknowledgments}
This work is supported by the National Science Foundation through the grant DMR-1708818. We would also like to thank Mark Dykman for thoughtful discussion on this work and related topics.



\bibliographystyle{elsarticle-num}
\bibliography{./csb_master}


\end{document}